\documentclass[10pt, conference, letterpaper]{IEEEtran}
\usepackage[letterpaper, left=0.65in, right=0.65in, bottom=1.01in, top=0.75in]{geometry}
\IEEEoverridecommandlockouts

\usepackage{cite}
\usepackage{amsmath,amssymb,amsfonts}
\usepackage{graphicx}
\usepackage[caption=false,font=footnotesize]{subfig}
\usepackage{textcomp}
\usepackage{xcolor}
\usepackage{tabularx}
\usepackage{multirow}
\usepackage{diagbox}
\usepackage[binary-units]{siunitx} 
\DeclareSIUnit{\bps}{bps}
\usepackage{flushend}
\usepackage{amsthm}
\def\tr{\mathrm{tr}}
\newtheorem{proposition}{Proposition}
\usepackage{lipsum}

\begin{document}

\title{Superimposed Transmission for Cooperative Cellular and Cell-Free Massive MIMO Systems\\
\thanks{This work was supported by the German Federal Ministry of Education and Research (BMBF) through \emph{Open6GHub+} (Grant no.  \emph{16KIS2402K}) project.}
}

\author{\IEEEauthorblockN{Wei Jiang}
\IEEEauthorblockA{\textit{Intelligent Networking Research Group}\\
\textit{German Research Center for Artificial Intelligence (DFKI)} \\
Kaiserslautern, 67663 Germany}
\and
\IEEEauthorblockN{Hans D. Schotten}
\IEEEauthorblockA{\textit{Department of
Electrical and Computer Engineering}\\
\textit{University of Kaiserslautern (RPTU)} \\
Kaiserslautern, 67663 Germany}
}

\maketitle

\begin{abstract}
This paper proposes a superimposed transmission strategy for cooperative cellular and cell-free massive MIMO systems. By classifying users into near and far, the base station transmits an additional data symbol for each near user, superimposed on the signals from distributed access points. Successive interference cancellation is employed at near-user receivers to decode both symbols. The proposed strategy achieves the highest peak spectral efficiency while maintaining fairness at the cell edge, thereby outperforming all the existing network configurations in system capacity.
\end{abstract}

\section{Introduction}

In conventional cellular networks, base stations (BSs) are typically located at the center of each cell, so users near the BS enjoy high quality-of-service (QoS) \cite{Ref_marzetta2010noncooperative}. In contrast, users at the cell edge suffer from severe performance degradation \cite{Ref_jiang2023cellfree}. The deployment of small cells in heterogeneous networks (HetNets)  \cite{Ref_cai2016green} can alleviate some dead zones, but it also increases interference, backhaul demand, and mobility-management complexity. 
Recently, cell-free (CF) massive MIMO (CFmMIMO) has been proposed to address cell-edge problems by employing many distributed access points (APs) that cooperatively serve users, resulting in uniformly high QoS \cite{Ref_ngo2017cellfree, Ref_jiang20226GCH9, Ref_nayebi2017precoding}. Later, user-centric (UC) approaches \cite{Ref_buzzi2020usercentric} have been proposed to enhance the efficiency of CF operation by dynamically clustering APs \cite{Ref_bjornson2020scalable}. However, a fully distributed architecture is often costly and sometimes impractical \cite{Ref_kim2022deployment}, as acquiring and wiring hundreds of AP sites significantly increases capital and operational costs \cite{Ref_jiang2024cost}  while facing deployment constraints. Furthermore, while such designs improve user fairness, the peak rates of well-conditioned users are severely restricted, which reduces system capacity.

Considering network operators’ large sunk investments in BS infrastructure, spectrum licenses, and backhaul networks, cooperative cellular and cell-free networks \cite{Ref_buzzi2024coexisting} have been explored that integrate CF elements with existing cellular networks \cite{Ref_kim2022deployment,Ref_zhang2024interdependent }. In this hierarchical massive MIMO (HmMIMO) paradigm \cite{Ref_jiang2024hierarchical, Ref_jiang2024heterogeneous}, legacy BSs act as the transceiver and also the central processor to coordinate clusters of distributed APs. This design reduces the number of AP sites and the scale of the fronthaul network, offering a cost-efficient approach to improving coverage at the network edge.

In this paper, we propose a superimposed transmission (ST) strategy to unlock the potential of cooperative cellular and cell-free networks. Alongside the data symbol transmitted by each user’s dedicated AP cluster, the BS selects a set of near users (NUs) and sends an \textit{additional} data symbol for each NU, superimposed on the cluster’s primary transmission. At the NU receivers, successive interference cancellation (SIC) is applied to first decode the BS’s symbol and then the AP’s signal. By layering an extra stream for NUs, the proposed scheme boosts peak rates of cell-center users while preserving fairness at the cell edge, thereby improving the overall system capacity.

This paper is structured as follows. Section II presents the system model. Section III details the superimposed scheme and spectral efficiency (SE) analysis. Section IV provides numerical results, and Section V concludes the work.

\section{System Model}

We consider the downlink of a cellular and cell-free cooperative network, as \cite{Ref_kim2022deployment,Ref_zhang2024interdependent, Ref_jiang2024hierarchical, Ref_jiang2024heterogeneous}, consists of one BS with \(N_b\) co-located antennas and \(L\) distributed APs, each with \(N_a\) antennas, where \(N_b \gg N_a\). The total number of antennas is \(M = N_b + LN_a\). The BS acts as a central transceiver and coordinates the APs via fronthaul links. Let \(\mathbb{L} = \{1,\ldots,L\}\) and \(\mathbb{K} = \{1,\ldots,K\}\) index the APs and users, respectively.
The channel between AP \(l\) and user \(k\) is \(\mathbf{h}_{kl} \in \mathbb{C}^{N_a}\), modeled as correlated Rayleigh fading:
\(
\mathbf{h}_{kl} \sim \mathcal{CN}(\mathbf{0}, \mathbf{R}_{kl})\),
where \(\mathbf{R}_{kl} = \mathbb{E}[\mathbf{h}_{kl} \mathbf{h}_{kl}^H]\) is the spatial correlation matrix, and its corresponding large-scale fading \(\beta_{kl} = \tr(\mathbf{R}_{kl}) / N_a\). Similarly, the channel between the BS and user \(k\) is \(\mathbf{h}_{k0} \in \mathbb{C}^{N_b}\), with \(\mathbf{h}_{k0} \sim \mathcal{CN}(\mathbf{0}, \mathbf{R}_{k0})\). Pilot contamination occurs when \(K > \tau_p\), where \(\tau_p\) is the length of orthogonal pilots. Using the linear minimum-mean-square-error (MMSE) estimation \cite{Ref_bjornson2020making}, the estimate of \(\mathbf{h}_{kl}\) follows the distribution $\hat{\mathbf{h}}_{kl} \sim \mathcal{CN}\left(\mathbf{0}, p_u\tau_p \mathbf{R}_{kl} \boldsymbol{\Gamma}_{kl}^{-1} \mathbf{R}_{kl} \right)$, where $\boldsymbol{\Gamma}_{kl} = p_u\tau_p \sum_{k' \in \mathcal{P}_k} \mathbf{R}_{k'l} + \sigma_n^2 \mathbf{I}_{N_a}$, \(p_u\) is the UE power constraint, and \(\mathcal{P}_k\) is the set of users sharing the same pilot as user \(k\). The estimation error \(\tilde{\mathbf{h}}_{kl} = \mathbf{h}_{kl} - \hat{\mathbf{h}}_{kl}\) has covariance matrix of
\(
\boldsymbol{\Theta}_{kl} = \mathbf{R}_{kl} - p_u \tau_p \mathbf{R}_{kl} \boldsymbol{\Gamma}_{kl}^{-1} \mathbf{R}_{kl}
\).
Thus, its distribution can be expressed by $\tilde{\mathbf{h}}_{kl} \sim \mathcal{CN}\left(\mathbf{0}, \boldsymbol{\Theta}_{kl} \right)$. For the BS channel, replacing \(l \to 0\), \(N_a \to N_b\), we get 
\(
\hat{\mathbf{h}}_{k0} \sim \mathcal{CN}(\mathbf{0}, p_u\tau_p \mathbf{R}_{k0} \boldsymbol{\Gamma}_{k0}^{-1} \mathbf{R}_{k0} )\) and
\(
\tilde{\mathbf{h}}_{k0} \sim \mathcal{CN}\left(\mathbf{0}, \boldsymbol{\Theta}_{k0} \right)
\), where  large-scale fading coefficients \(\beta_{k0} = \tr(\mathbf{R}_{k0}) / N_b\).

\section{Downlink Superimposed Transmission}
This section details the communication process, including user classification, ST at the transmitter, and SIC at the receiver, with achievable SE derivation. 

\subsection{User Classification and User-Centric Clustering}
The process begins by classifying users into two groups: NUs and far users (FUs), based on slowly-varying metrics, which are assumed known \textit{a priori}. For example, the set of NUs can be formed as $\mathbb{K}_0 = \{k : \beta_{k0} \geqslant \bar{\beta}_0\}$, where $\bar{\beta}_0$ stands for a threshold. Alternatively, a geometric criterion can be used, i.e.,  $\mathbb{K}_0 = \{k : d_k \leqslant d_0\}$ where $d_k$ is the user-BS distance and $d_0$ is a threshold radius. Users not in $\mathbb{K}_0$ are designated as FUs. Subsequently, a user-centric cluster of APs, denoted $\mathbb{L}_k$, is determined for each user $k \in \mathbb{K}$, following \cite{Ref_buzzi2020usercentric, Ref_bjornson2020scalable}. Conversely, each AP $l$ maintains a set of associated users defined as $\mathbb{K}_l = \{k : l \in \mathbb{L}_k\}$. The choice of these thresholds governs the NU/FU split and the cluster sizes, creating a design parameter.

\subsection{Superimposed Signal Transmission}

In this approach, the APs transmit $K$ zero-mean, unit-variance data symbols $\{x_{k,2}\}_{k\in \mathbb{K}}$ to $K$ users. These symbols are mutually uncorrelated, i.e., $\mathbb{E}[x_{k,2}x_{k',2}^*]=0$ for $k'\neq k$ and $\mathbb{E}[|x_{k,2}|^2]=1$. Let $\mathbf{w}_{kl}\in \mathbb{C}^{N_a}$ be the precoding vector for user $k$ at AP $l$. The transmitted signal from AP $l$ is $\mathbf{s}_l = \sqrt{p_a}\sum\nolimits_{k\in \mathbb{K}_l} \sqrt{\eta_{kl}}\mathbf{w}_{kl} x_{k,2}$, where $p_a$ is the maximum AP transmit power and $\eta_{kl}\in [0,1]$ is the power coefficient for user $k$ at AP $l$. The per-AP power constraint $\mathbb{E}[\|\mathbf{s}_l\|^2]\leqslant p_a$ requires that $\sum_{k\in \mathbb{K}_l} \eta_{kl}\leqslant 1$. 
A key feature of our approach is that the BS simultaneously superimposes an additional symbol $x_{k,1}$ for each NU $k \in \mathbb{K}_0$. The BS transmits $\mathbf{s}_0 = \sqrt{p_b}\sum\nolimits_{k\in \mathbb{K}_0} \sqrt{\eta_{k0}} \mathbf{w}_{k0} x_{k,1}$, where $\mathbf{w}_{k0}\in \mathbb{C}^{N_b}$ denote the precoding vector for user $k$, normalized such that $\mathbb{E}[\|\mathbf{w}_{k0}\|^2]= 1$, $p_b$ is the maximum BS transmit power, and $\eta_{k0}\in [0,1]$ is the power allocation coefficient for user $k$, satisfying $\sum_{k\in \mathbb{K}_0} \eta_{k0}\leqslant 1$. 

As a result, user $k$ observes  $y_k  =  \mathbf{h}_{k0}^T\mathbf{s}_0 +  \sum\nolimits_{l\in \mathbb{L} } \mathbf{h}_{kl}^T\mathbf{s}_l  +n_k  $, corrupted by additive noise $n_k\sim \mathcal{CN}(0,\sigma^2_n)$.
Substituting the expressions for $\mathbf{s}_0$ and $\mathbf{s}_l$ yields the expanded received signal model:
\begin{align}   \label{eQn_downlinkModel} 
   y_k =  & \sqrt{p_b}\sum\limits_{k'\in \mathbb{K}_0} \sqrt{\eta_{k'0}} \mathbf{h}_{k0}^T \mathbf{w}_{k'0} x_{k',1}\\  \nonumber
     + & \sqrt{p_a} \sum\limits_{l\in\mathbb{L}} \sum\limits_{k'\in \mathbb{K}_l}  \sqrt{\eta_{k'l}} \mathbf{h}_{kl}^T \mathbf{w}_{k'l} x_{k',2} +n_k.
\end{align}
Among commonly used schemes, the MMSE precoding optimally balances interference suppression with noise amplification. A variant known as \textit{local partial} MMSE (LP-MMSE) \cite{Ref_bjornson2020scalable}, is well tailored for user-centric clustering.  For example, the normalized LP-MMSE precoder for AP $l$ is given by
\begin{equation} \label{GS_LPMMSE}
\mathbf{w}_{kl}= \frac{\left[\left( \sum\limits_{k\in \mathbb{K}_l } p_u \left(\hat{\mathbf{h}}_{kl}\hat{\mathbf{h}}_{kl}^H+ \boldsymbol{\Theta}_{kl} \right ) + \sigma_n^2 \mathbf{I}_{N_t} \right)^{-1} \hat{\mathbf{h}}_{kl}\right]^*} {\left\|\left( \sum\limits_{k\in \mathbb{K}_l } p_u \left(\hat{\mathbf{h}}_{kl}\hat{\mathbf{h}}_{kl}^H+ \boldsymbol{\Theta}_{kl} \right ) + \sigma_n^2 \mathbf{I}_{N_t} \right)^{-1} \hat{\mathbf{h}}_{kl}\right\|}.    
\end{equation}  
Note that the proposed scheme is general for any precoder, like zero-forcing \cite{Ref_interdonato2020local, Ref_nayebi2017precoding}  or conjugate beamforming \cite{Ref_ngo2017cellfree}.

\subsection{SIC Detection and Achievable SE for Near Users}

For an NU $k$, the received signal contains a superimposition of $x_{k,1}$ and $x_{k,2}$. Different from \cite{Ref_ding2017survey, Ref_Bashar2019}, which rely on power allocation for non-orthogonal multiple access, the proposed scheme exploits the inherent geometric advantage, where the BS naturally provides a stronger signal to an NU, facilitating the separation of the superimposed signal. Hence, the receiver first decodes $x_{k,1}$ by treating all signals from the APs as interference. After decoding $x_{k,1}$, the receiver performs SIC to detect $x_{k,2}$. Massive MIMO uses TDD reciprocity to avoid prohibitive overhead of downlink pilots that scales with $M$, meaning user $k$ cannot know $\{\hat{\mathbf{h}}_{kl}\}_{l \in \mathbb{L}_0}$. This user must therefore detect $y_k$ non-coherently using the statistical expectations, i.e., $\{ \mathbb{E} [ \mathbf{h}_{kl}^T \mathbf{w}_{kl} ] \}_{l \in \mathbb{L}_0}$. 
Consequently, \eqref{eQn_downlinkModel} is decomposed into  
\begin{align} \nonumber \label{eQn_DLGeneralSig}
    y_k = &  \underbrace{   \sqrt{p_b\eta_{k0}} \mathbb{E} [   \mathbf{h}_{k0}^T \mathbf{w}_{k0}  ] x_{k,1}}_{\mathcal{S}_1:\:\text{Desired signal}}  \\ + & \underbrace{ \sqrt{p_b\eta_{k0}}\left(\mathbf{h}_{k0}^T \mathbf{w}_{k0} -\mathbb{E} [   \mathbf{h}_{k0}^T \mathbf{w}_{k0}  ]\right)x_{k,1}    }_{\mathcal{J}_1:\:\text{Channel\:uncertainty\:error}} \\  + &
    \underbrace{  \left ( \begin{aligned} &\sqrt{p_b}\sum\limits_{k'\in \mathbb{K}_0 \backslash \{k\}} \sqrt{\eta_{k'0}}\mathbf{h}_{k0}^T \mathbf{w}_{{k'0}}  x_{k',1}  + \\ \nonumber
    &\sqrt{p_a}\sum\limits_{l\in \mathbb{L}}\sum\limits_{k'\in \mathbb{K}_l } \sqrt{\eta_{k'l}} \mathbf{h}_{kl}^T  \mathbf{w}_{{k'l}} x_{k',2} \end{aligned}  \right) 
    }_{\mathcal{J}_2:\:\text{Inter-user\:interference}}+ n_k. 
\end{align}

Using the standard capacity lower bounds (cf. \cite[Prop. 3]{Ref_bjornson2020scalable}), the downlink signal-to-interference-plus-noise ratio (SINR) for detecting $x_{k,1}$ is given by 
\begin{align}  \label{EQN_DL_sinr_fP}
    \gamma_{k,1}  = \frac{|\mathcal{S}_1|^2}{\mathbb{E}\left[|\mathcal{J}_1+\mathcal{J}_2|^2\right] + \sigma_n^2},
\end{align}
where $|\mathcal{S}_1|^2 =  | \sqrt{p_b\eta_{k0}} \mathbb{E} [   \mathbf{h}_{k0}^T \mathbf{w}_{k0}  ]  |^2$.
Given the independence of the data symbols, i.e., $\mathbb{E}[x_{k}^*x_{k'}]=0$ for $k'\neq k$, the terms $\mathcal{J}_1$ and $\mathcal{J}_2$ in \eqref{eQn_DLGeneralSig} are uncorrelated. This implies $\mathbb{E}\left[|\mathcal{J}_1 + \mathcal{J}_2|^2\right]=\mathbb{E}\left[|\mathcal{J}_1|^2\right]+\mathbb{E}\left[|\mathcal{J}_2|^2\right]$. The variance of $\mathcal{J}_1$ is computed as
\begin{align}  \label{APPEQ1}
    \mathbb{E}\left[|\mathcal{J}_1|^2\right]  =   p_b\eta_{k0} \mathbb{E}\left[ \left| \mathbf{h}_{k0}^T \mathbf{w}_{k0} -\mathbb{E} [   \mathbf{h}_{k0}^T \mathbf{w}_{k0}  ] \right|^2 \right],
\end{align}
which is further derived as
\begin{align} \label{GS_T1inDL}
    \mathbb{E}\left[|\mathcal{J}_1|^2\right] =  p_b\eta_{k0} \left(\mathbb{E}\left[ \left| \mathbf{h}_{k0}^T \mathbf{w}_{k0}\right|^2 \right] - \left|\mathbb{E}[ \mathbf{h}_{k0}^T \mathbf{w}_{k0} ] \right|^2 \right).
\end{align}
Next, the variance of $\mathcal{J}_2$ is given by
\begin{align} \label{GS_VarianceJ2} \nonumber
    \mathbb{E}\left[|\mathcal{J}_2|^2\right]= &  p_b\sum\limits_{k'\in \mathbb{K}_0 \backslash \{k\}}  \eta_{k'0}   \mathbb{E}[| \mathbf{h}_{k0}^T  \mathbf{w}_{{k'0}}    |^2]  \\
     + & p_a  \sum\limits_{l\in\mathbb{L}} \sum\limits_{k'\in \mathbb{K}_l }   \eta_{k'l} \mathbb{E}[|  \mathbf{h}_{kl}^T  \mathbf{w}_{{k'l}}   |^2].
\end{align}
Substituting \eqref{GS_T1inDL} and \eqref{GS_VarianceJ2} into \eqref{EQN_DL_sinr_fP} yields the effective SINR for detecting $x_{k,1}$:
\begin{equation} \label{eQn:DLSINR1}
    \gamma_{k,1} = \frac{ p_b\eta_{k0} \left|  \mathbb{E} [   \mathbf{h}_{k0}^T \mathbf{w}_{k0}  ] \right|^2 }{  \left\{ \begin{aligned}
     &p_b \sum\limits_{k'\in \mathbb{K}_0 } \eta_{k'0}   \mathbb{E}[| \mathbf{h}_{k0}^T  \mathbf{w}_{{k'0}}    |^2] {-} p_b\eta_{k0}  \left|\mathbb{E}[ \mathbf{h}_{k0}^T \mathbf{w}_{k0} ] \right|^2  \\
     &+  p_a  \sum\limits_{l\in\mathbb{L}} \sum\limits_{k'\in \mathbb{K}_l  }   \eta_{k'l} \mathbb{E}[|  \mathbf{h}_{kl}^T  \mathbf{w}_{{k'l}}   |^2] +\sigma_n^2 \end{aligned} \right\}   }.
\end{equation}

Assuming the SINR $\gamma_{k,1}$ is sufficient for $x_{k,1}$ to be successfully detected ($\hat{x}_{k,1}=x_{k,1}$), its component is subtracted from $y_k$. The residual signal $y'_k = y_k - \sqrt{p_b\eta_{k0}}  \mathbf{h}_{k0}^T \mathbf{w}_{k0}  x_{k,1}$ is used to detect $x_{k,2}$ and can be expanded as:
\begin{align} \nonumber \label{eQn_DL_residualSignal}
    y'_k = &  \underbrace{  \sqrt{p_a} \sum\limits_{l\in \mathbb{L}_k} \sqrt{\eta_{kl}} \mathbb{E} [\mathbf{h}_{kl}^T  \mathbf{w}_{{kl}}] x_{k,2}    }_{\mathcal{S}_2:\:\text{Desired Signal}}  \\ + & \underbrace{ \sqrt{p_a} \sum\limits_{l\in \mathbb{L}_k} \sqrt{\eta_{kl}}\left(\mathbf{h}_{kl}^T \mathbf{w}_{kl} -\mathbb{E} [   \mathbf{h}_{kl}^T \mathbf{w}_{kl}  ]\right)x_{k,2}    }_{\mathcal{J}_{3}:\:\text{Channel Uncertainty Error}} \\  + &
    \underbrace{  \left ( \begin{aligned} &\sqrt{p_b}\sum\limits_{k'\in \mathbb{K}_0 \backslash \{k\}} \sqrt{\eta_{k'0}}\mathbf{h}_{k0}^T \mathbf{w}_{{k'0}}  x_{k',1}  + \\ \nonumber
    &\sqrt{p_a}\sum\limits_{l\in \mathbb{L}}\sum\limits_{k'\in \mathbb{K}_l \backslash \{k\} } \sqrt{\eta_{k'l}} \mathbf{h}_{kl}^T  \mathbf{w}_{{k'l}} x_{k',2} \end{aligned}  \right)
    }_{\mathcal{J}_{4}:\:\text{Inter-User Interference}}+ n_k.
\end{align}
Following a similar derivation to that of $\gamma_{k,1}$, the SINR for detecting $x_{k,2}$ is
\begin{equation} \label{eQn:DLSINR2}
    \gamma_{k,2}= \frac{  p_a\left|  \sum_{l\in \mathbb{L}_k } \sqrt{\eta_{kl}}\mathbb{E} [\mathbf{h}_{kl}^T\mathbf{w}_{kl}] \right|^2  }{
    \left\{ \begin{aligned}
      p_b&\sum\limits_{k'\in \mathbb{K}_0 \backslash \{k\} }  \eta_{k'0}   \mathbb{E}[| \mathbf{h}_{k0}^T  \mathbf{w}_{{k'0}}    |^2] \\
     & +  p_a  \sum\limits_{l\in\mathbb{L}} \sum\limits_{k'\in \mathbb{K}_l  }   \eta_{k'l} \mathbb{E}[|  \mathbf{h}_{kl}^T  \mathbf{w}_{{k'l}}   |^2] \\
     & - p_a \sum\limits_{l \in \mathbb{L}_k}  \eta_{kl}  \left|\mathbb{E}[ \mathbf{h}_{kl}^T \mathbf{w}_{kl} ] \right|^2  +\sigma_n^2 \end{aligned} \right\}
    }.
\end{equation}

\begin{proposition}
The achievable spectral efficiency for a near user $k\in \mathbb{K}_0$, receiving the data symbol pair $(x_{k,1}, x_{k,2})$, is given by $R_{k}=\log_2(1+\gamma_{k,1})+\log_2(1+\gamma_{k,2})$, where $\gamma_{k,1}$ and $\gamma_{k,1}$ are given by \eqref{eQn:DLSINR1} and \eqref{eQn:DLSINR2}, respectively.
\end{proposition}

\subsection{Detection and Achievable SE for Far Users}

For any FU $k$, its data symbol $x_{k,2}$ is transmitted solely from its serving user-centric APs.  Consequently, this user directly detects $x_{k,2}$ without SIC, where the received signal  is decomposed as follows:
\begin{align} \nonumber \label{eQn_DL_desiredSignal_farUser}
    y_k = &  \underbrace{  \sqrt{p_a} \sum\limits_{l\in \mathbb{L}_k} \sqrt{\eta_{kl}} \mathbb{E} [\mathbf{h}_{kl}^T  \mathbf{w}_{{kl}}] x_{k,2}    }_{\mathcal{S}_3:\:\text{Desired Signal}}  \\ + & \underbrace{ \sqrt{p_a} \sum\limits_{l\in \mathbb{L}_k} \sqrt{\eta_{kl}}\left(\mathbf{h}_{kl}^T \mathbf{w}_{kl} -\mathbb{E} [   \mathbf{h}_{kl}^T \mathbf{w}_{kl}  ]\right)x_{k,2}    }_{\mathcal{J}_5:\:\text{Channel Uncertainty Error}} \\ \nonumber + &
    \underbrace{  \left ( \begin{aligned} &\sqrt{p_b}\sum\limits_{k'\in \mathbb{K}_0 } \sqrt{\eta_{k'0}}\mathbf{h}_{k0}^T \mathbf{w}_{{k'0}}  x_{k',1}  + \\
    &\sqrt{p_a}\sum\limits_{l\in \mathbb{L}}\sum\limits_{k'\in \mathbb{K}_l \backslash \{k\} } \sqrt{\eta_{k'l}} \mathbf{h}_{kl}^T  \mathbf{w}_{{k'l}} x_{k',2} \end{aligned}  \right)
    }_{\mathcal{J}_6:\:\text{Inter-User Interference}}+ n_k.
\end{align}
\begin{proposition}
The achievable spectral efficiency for a far user $k\notin \mathbb{K}_0$ is given by $R_{k}= \log_2(1 + \xi_{k})$, where the instantaneous effective SINR is expressed as
\begin{equation} \label{eQn:DLSINR_faruser_01}
    \xi_{k}= \frac{  p_a\left|  \sum_{l\in \mathbb{L}_k } \sqrt{\eta_{kl}}\mathbb{E} [\mathbf{h}_{kl}^T\mathbf{w}_{kl}] \right|^2  }{
    \left\{ \begin{aligned}
     p_b&\sum\limits_{k'\in \mathbb{K}_0  } \eta_{k'0}   \mathbb{E}[| \mathbf{h}_{k0}^T  \mathbf{w}_{{k'0}}    |^2]  \\
     &+ p_a  \sum\limits_{l\in\mathbb{L}} \sum\limits_{k'\in \mathbb{K}_l  }   \eta_{k'l} \mathbb{E}[|  \mathbf{h}_{kl}^T  \mathbf{w}_{{k'l}}   |^2] \\
     &- p_a \sum\limits_{l \in \mathbb{L}_k}  \eta_{kl}  \left|\mathbb{E}[ \mathbf{h}_{kl}^T \mathbf{w}_{kl} ] \right|^2  +\sigma_n^2
    \end{aligned} \right\}
    }.
\end{equation}
\end{proposition}

\begin{figure*}[!tbph]
\centerline{ 
\subfloat[]{
\includegraphics[width=0.45\textwidth]{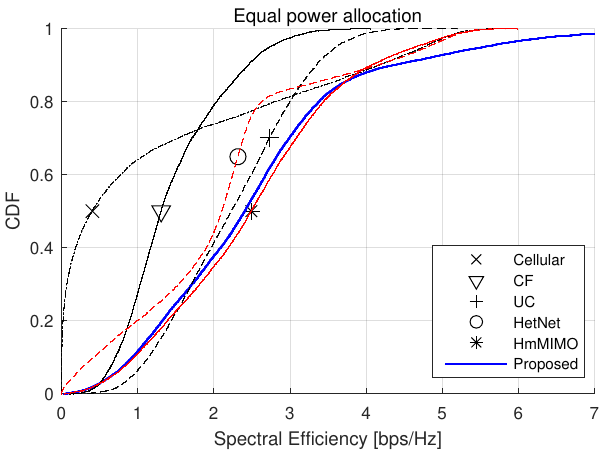}
\label{fig:epa}
}
\hspace{20mm}
\subfloat[]{
\includegraphics[width=0.45\textwidth]{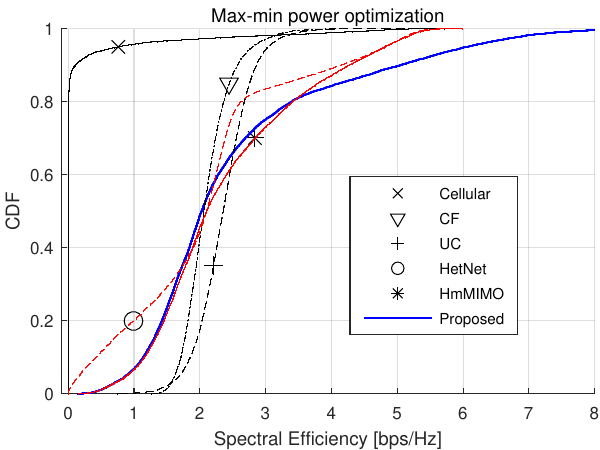}
\label{fig:mm}
}
}
\hspace{15mm}
\caption{CDF of per-user SE under equal power allocation (a)  and  max-min power control (b).}
\label{Fig_performance}
\end{figure*}

\begin{figure}[!tbph]
\centerline{ 
\subfloat[]{
\includegraphics[width=0.45\textwidth]{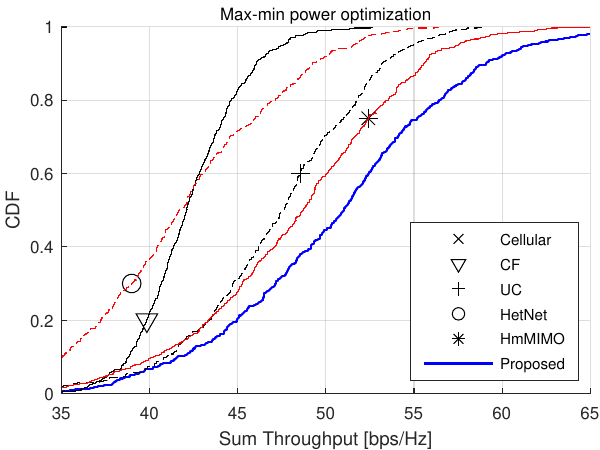}
}
}
\hspace{15mm}
\caption{CDF of sum throughput under  max-min power control.}
\label{fig:sum}
\end{figure}

\begin{figure*}[!tbph]
\centerline{ 
\subfloat[]{
\includegraphics[width=0.45\textwidth]{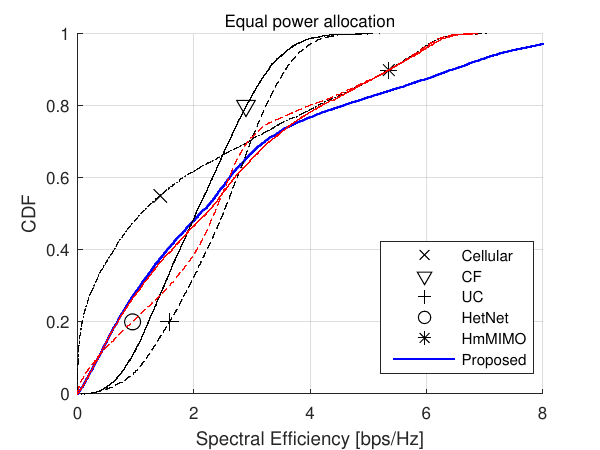}
\label{fig:epaK20}
}
\hspace{20mm}
\subfloat[]{
\includegraphics[width=0.45\textwidth]{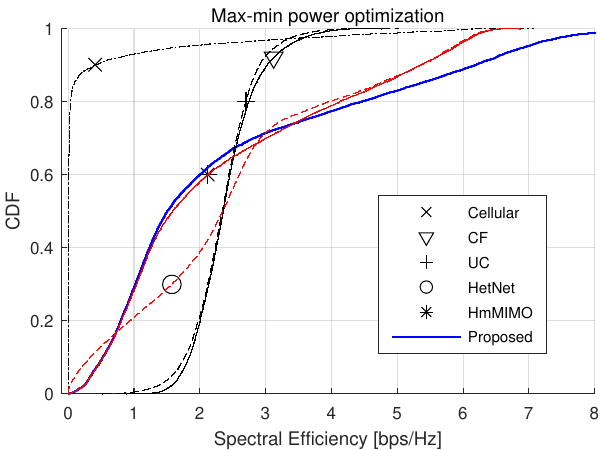}
\label{fig:mmK20}
}
}
\hspace{15mm}
\caption{CDF of per-user SE under equal power allocation (a)  and  max-min power control (b) when the number of users is increased to 20.}
\label{Fig_performanceK20}
\end{figure*}

\section{Numerical Evaluation}

This section provides performance evaluation of the proposed scheme against several benchmarks. The simulation parameters and benchmark configurations are detailed below: a total of \(M=120\) antennas are deployed to serve \(K=10\) active users (by default) uniformly distributed within a circular coverage area of radius $1$ km. Each coherence-block length lasts \(200\) symbols and the pilot sequence length is \(\tau_p=5<K\), intentionally induces pilot contamination.  Large-scale fading follows the channel model in \cite{Ref_jiang2021cellfree}, and small-scale spatial correlation is incorporated using the Gaussian  scattering  \cite[Sec.~2.6]{SIG-093} for uniform linear arrays, with angular standard deviation of $10^\circ$.
Each service antenna is subject to a power budget of \(50\ \mathrm{mW}\); accordingly, a BS with 32 antennas can transmit up to \(1.6\ \mathrm{W}\), while a four-antenna AP is limited to \(200\ \mathrm{mW}\) transmit power. The maximal transmit power for user equipment is set to \(p_u=200\ \mathrm{mW}\).  The system operates with a noise power spectral density of \(-174\ \mathrm{dBm/Hz}\), a noise figure of \(9\ \mathrm{dB}\), and a bandwidth of \(5\ \mathrm{MHz}\). 

In our simulation, the cooperative network comprises a BS with $32$ antennas located at the center of the coverage area, plus $24$ randomly distributed APs, each equipped with $N_a=4$ antennas. Users within 400 meters of the BS are classified as NUs, while 8 closest APs for each user form its user-centric cluster. 
For fair and informative comparisons, we consider five benchmarks:  
\begin{enumerate}
    \item \textbf{HmMIMO: } utilizes the same network topology (one 32-antenna BS and 24 four-antenna APs) with \emph{non-superimposed} precoding \cite{Ref_jiang2024hierarchical};
    \item \textbf{CF}: 32 four-antenna APs jointly serve all users in a cell-free fashion \cite{Ref_ngo2017cellfree};
    \item \textbf{UC:}  same AP layout as CF but each user is served by its 16 closest APs \cite{Ref_buzzi2020usercentric} (totaling $64$ service antennas) to ensure a fair comparison;
    \item  \textbf{Cellular} massive MIMO: a traditional system with all 128 antennas at the BS (no distributed APs) \cite{Ref_marzetta2010noncooperative};
    \item \textbf{HetNet:} 32-antenna macro BS and 24 small-cell micro BSs (4 antennas each). The macro and micro tiers independently serve disjoint user sets, with no BS cooperation \cite{Ref_cai2016green}.
\end{enumerate}

Performance is compared using the empirical cumulative distribution function (CDF) of per-user SE and sum throughput. We implement LP-MMSE precoding in \eqref{GS_LPMMSE} under both equal power allocation (EPA) and max-min power control. In such CDF curves, the lower tail represents the worst-case performance and the $5^{th}$-percentile SE is a standard metric for assessing user fairness. Conversely, the upper tail reflects the peak rates achievable by users with favorable channel conditions. 
Centralized architectures (Cellular and HetNet), as shown in \figurename~\ref{fig:epa}, achieve high peak rates but suffer from very low $5^{th}$-percentile SE and poor fairness. Fully distributed systems (CF and UC), on the other hand, enhance fairness but compromise on peak rates. The HmMIMO baseline strikes a balance, combining the high peak rates of centralized systems with the improved fairness of distributed ones. With the aid of ST and SIC, the proposed scheme boosts the peak rates of the HmMIMO baseline without sacrificing user fairness. The figure shows that it achieves the best performance across nearly the entire SE distribution.  It matches the $5^{th}$-percentile SE of HmMIMO—surpassed only by UC, which employs twice the number of distributed antennas—while significantly outperforming all other schemes, including CF.  In the upper tail, the proposed scheme demonstrates clear performance superiority with a $95^{th}$-percentile SE of approximately \SI{5.6}{bps/Hz}. This represents a \SI{1.0}{bps/Hz} (improving 21.7\%) gain over HmMIMO and centralized approaches, and substantially higher than the distributed schemes—roughly 155\% and 207\% of those achieved by UC (\SI{3.6}{bps/Hz}) and CF (\SI{2.7}{bps/Hz}).

Max–min power allocation is commonly employed in CF systems to effectively improve user fairness. As shown in \figurename~\ref{fig:mm}, this strategy enables fully distributed schemes (CF and UC) to attain superior $5^{th}$-percentile SE. HmMIMO also maintains good user fairness, substantially outperforming the purely centralized schemes. Notably, the proposed scheme preserves user fairness relative to the HmMIMO baseline while simultaneously boosting its peak rate. Specifically, at the 95th percentile, it reaches \SI{6.1}{bps/Hz}, outperforming HmMIMO (\SI{4.8}{bps/Hz}) by \SI{1.3}{bps/Hz}—an improvement of about $27.1\%$. This performance significantly exceeds CF (\SI{3.0}{bps/Hz}) by \SI{3.1}{bps/Hz}—i.e., more than double CF’s peak. Furthermore, the proposed scheme yields system-capacity gains. As \figurename~\ref{fig:sum} demonstrates, its sum throughput under max–min power control attains an average ($50^{th}$-percentile) of \SI{51}{bps/Hz}, an increase of around \SI{2.3}{bps/Hz} ($4.7\%$) over HmMIMO’s \SI{48.7}{bps/Hz} and substantially higher than the other benchmarks. To validate the generality of the results, we repeated the simulation with the number of users increased to $K=20$. As shown in \figurename \ref{Fig_performanceK20}, a similar performance gain of the proposed scheme is observed.

\section{Conclusions}
This paper proposed a superimposed transmission strategy for cooperative cellular and cell-free systems. By transmitting an additional data stream from the base station to near users and applying successive interference cancellation at the receiver, it significantly enhances spectral efficiency while maintaining cell-edge fairness. Numerical results demonstrate a peak spectral efficiency of \SI{6.1}{bps/Hz}—a $27.1\%$ gain over centralized/hierarchical massive MIMO and more than $200\%$ over distributed cell-free systems. It offers a practical and backward-compatible strategy for enhancing capacity towards next-generation wireless networks.


\bibliographystyle{IEEEtran}
\bibliography{IEEEabrv,Ref_COML}

\end{document}